# Zero Trust: Applications, Challenges, and Opportunities

[a] Saeid Ghasemshirazi, [b] Ghazaleh Shirvani [c] Mohammad Ali Alipour

a   Department of Industrial Engineering, Iran University of Science and Technology, Tehran, Iran, saeidgs@yahoo.com

b   Department of Computer Engineering, Iran University of Science and Technology, Tehran, Iran, ghazaleh.sh3p@gmail.com

C   Department of Energy Management and Optimization, Institute of Science and High Technology and Environmental Sciences, Graduate University of Advanced Technology, Kerman, Iran, m.alialipour77@gmail.com

**Abstract**

The escalating complexity of cybersecurity threats necessitates innovative approaches to safeguard digital assets and sensitive information. The Zero Trust paradigm offers a transformative solution by challenging conventional security models and emphasizing continuous verification and least privilege access. This survey comprehensively explores the theoretical foundations, practical implementations, applications, challenges, and future trends of Zero Trust. Through meticulous analysis, we highlight the relevance of Zero Trust in securing cloud environments, facilitating remote work, and protecting the Internet of Things (IoT) ecosystem. While cultural barriers and technical complexities present challenges, their mitigation unlocks Zero Trust's potential. Integrating Zero Trust with emerging technologies like AI and machine learning augments its efficacy, promising a dynamic and responsive security landscape. Embracing Zero Trust empowers organizations to navigate the ever-evolving cybersecurity realm with resilience and adaptability, redefining trust in the digital age.

# 1. Introduction

In the rapidly evolving landscape of information technology and network security, traditional perimeter-based security models have proven inadequate in safeguarding digital assets from an ever-expanding array of sophisticated cyber threats. Consequently, a paradigm shift in security strategies has emerged, giving rise to the innovative concept of "Zero Trust." Rooted in the principles of constant verification, least privilege access, and strict segmentation, Zero Trust offers a modern approach to protect digital infrastructures in the face of dynamic and pervasive security challenges (Cao et al., 2022). Once grounded on the belief that an organization's internal network could be implicitly trusted, the traditional security model has become increasingly vulnerable as cyber attackers persistently exploit various vulnerabilities. Legacy security measures, such as firewalls and VPNs, have struggled to cope with the complexities of cloud computing, themobile workforce, and the burgeoning Internet of Things (IoT) ecosystem. The reliance on static perimeters and "castle-and-moat" security strategies have proven ineffective against advanced threats that infiltrate organizations and linger undetected (Egerton et al., 2021). ,On the other hand Zero Trustembraces an approach that defies the conventional wisdom of implicit trust, rejecting the notion of a trusted internal network. Under the Zero Trust framework, each user, device, and application, regardless of its location or origin, is deemed untrusted until it proves its authenticity and authorization through continuous verification. By assuming a "never trust, always verify" mantra, Zero Trust upends the traditional trust assumptions and necessitates ongoing authentication, authorization, and validation of entities attempting to access critical resources (Alevizos et al., 2022). Central to the Zero Trust philosophy is the principle of least privilege, whereby access permissions are tightly controlled, granting users minimal privileges essential to their roles and functions. This granularity ensures that potential threats are contained and mitigated, preventing lateral movement within the network. Furthermore, Zero Trust advocates for rigorous network segmentation, dividing the infrastructure into smaller, isolated zones that act as virtual micro-perimeters, restricting thelateral traversal of adversaries (Buck et al., 2021a). The applications of Zero Trust span across diverse industries and organizations, including government agencies, financial institutions, healthcare providers, and enterprises of varying sizes. The architecture's adaptability and scalability make it a compelling choice for modern networks that demand flexibility, dynamic access controls, and resilient security postures. Despite its potential benefits, the implementation of Zero Trust is challenging. Organizations often encounter resistance to change, cultural barriers, and difficulties transitioning from legacy systems to Zero Trust architectures. Additionally, technical complexities and integration issues may arise when aligning diverse technologies and platforms within the organization. Understanding these challenges is critical to devising effective strategies for successful Zero Trust deployment (Astillo et al., 2021). This paper presents a comprehensive survey examining the applications, challenges, and opportunities associated with Zero Trust. We aim to provide a nuanced understanding of how Zero Trust is being adopted across industries, its practical use cases, and the hurdles faced during implementation (Astillo et al., 2021). Moreover, we explore the cybersecurity and privacy considerations that arise with the adoption of Zero Trust, shedding light on its role in countering emerging threats. By analyzing existing literature and collecting data from IT professionals, cybersecurity experts, and network administrators, we seek to contribute to the ongoing discourse on Zero Trust, offering valuable insights and practical recommendations for organizations looking to bolster their security posture in the digital age. Through this exploration, we aim to inspire further innovation in network security and foster a safer, more resilient cyberspace for all stakeholders involved (Alagappan et al., 2022).

| Abbreviations | Definitions |
| --- | --- |
| ZT | Zero Trust |
| ZTN | Zero Trust Network |
| ZTE | Zero Trust Edge |
| NIST | National Institute of Standards and Technology |
| IoT | Internet of Things |
| VPN | Virtual Private Network |
| BYOD | Bring Your Device |
| SASE | Secure Access Service Edge |
| IAM | Identity and Access Management |
| API | Application Programming Interface |
| SD-WAN | Software-Defined Wide Area Network |
| MFA | Multi-Factor Authentication |
| EDR | Endpoint Detection and Response |
| SOC | Security Operations Center |

Table 1. Abbreviations and Definitions Used in the Paper

This survey aims to investigate the various applications, challenges, and opportunities associated with the implementation of Zero Trust architecture in modern organizations. By exploring the experiences and perspectives of IT professionals, cybersecurity experts, and other relevant stakeholders, this study seeks to provide valuable insights into the current state of Zero Trust adoption, its impact on organizational security, and its potential benefits.

## 2. Background and Motivation

In recent years, the digital landscape has witnessed a rapid proliferation of cyber threats and sophisticated attacks that have compromised the security of organizations across the globe. Traditional security models, centered around perimeter-based defenses and implicit trust, need to be revised in safeguarding critical assets and sensitive information. The emergence of cloud computing, mobile workforces, and the Internet of Things (IoT) has further challenged the efficacy of conventional security strategies, leaving organizations vulnerable to evolving threats (Rahmadika et al., 2023). In response to this escalating threat landscape and the need for more robust and dynamic security solutions, the concept of Zero Trust has gained traction in network security. Initially proposed by Forrester Research in 2010, the Zero Trust model aims to address the limitations of traditional perimeter-centric security approaches by fundamentally altering how trust is established and verified within a network (Teerakanok et al., 2021a). The motivation behind this paper lies in the urgency to explore and understand the applications, challenges, and opportunities presented by the Zero Trust paradigm. As organizations increasingly embrace digital transformation and adopt new technologies, resilient and adaptive security measures become paramount. By shifting from a trust-based approach to continuous verification and strict access controls, Zero Trust promises to bolster network security and mitigate a wide range of cyber threats (Velumani et al., 2022). Through this research, we aim to analyse Zero Trust's modern concepts, practical implications, and relevance in safeguarding organizations against contemporary cyber threats. By identifying its applications across industries and assessing the challenges faced during implementation, we seek to offer valuable insights to network administrators, IT professionals, and cybersecurity experts looking to fortify their defences (Ramezanpour and Jagannath, 2022). Moreover, this paper aspires to contribute to the broader cybersecurity community and inform policymakers and decision-makers about the merits of adopting Zero Trust as a proactive defence mechanism. By understanding the potential benefits and limitations of Zero Trust, organizations can make informed decisions and devise robust security strategies that align with their specific requirements and risk profiles (Gupta et al., 2018). As cyber threats continue to evolve and exploit vulnerabilities in traditional security models, this research endeavors to shed light on the transformative potential of Zero Trust. By fostering a deeper understanding of this innovative security concept, we aim to empower organizations to embrace Zero Trust principles and fortify their networks against the digital age's relentless and ever-evolving cyber threats (Ray, 2023).

## 3. Research Objectives:

The primary purpose of this survey is to comprehensively investigate the applications, challenges, and opportunities associated with implementing the Zero Trust (ZT) paradigm in modern network security. By gathering insights from IT professionals, cybersecurity experts, and network administrators, we aim to achieve the following specific research objectives:

### 3.1. Assess the Current State of Zero Trust Adoption:

We seek to gauge Zero Trust adoption across different industries and organizations. By understanding the extent to which organizations have embraced Zero Trust principles, we can identify trends and patterns that highlight its relevance in today's cybersecurity landscape.

### 3.2. Identify Key Applications of Zero Trust:

Through this survey, we aim to uncover the diverse applications of Zero Trust architecture in various domains. By examining real-world use cases and success stories, we aim to showcase how Zero Trust is effectively deployed to protect critical assets and secure sensitive data.

3.3. **Explore Challenges Faced during Zero Trust Implementation:**

Understanding the obstacles and difficulties encountered during Zero Trust deployment is crucial to devising effective strategies for successful implementation. We aim to identify organizational, technical, and financial challenges that may hinder the adoption of Zero Trust and propose potential solutions.

3.4. **Evaluate Zero Trust's Effectiveness in Addressing Cybersecurity Threats:**

This survey assesses how Zero Trust measuresagainst contemporary cyber threats. By analyzing the experiences and perspectives of survey participants, we aim to ascertain the efficacy of Zero Trust in mitigating various attack vectors.

3.5. **Investigate Privacy and Data Protection Considerations:**

As Zero Trust involves continuous verification and access controls, we will explore the implications for data privacy and regulatory compliance. By examining the intersection of Zero Trust and privacy, we aim to provide a comprehensive view of its impact on data protection.

3.6. **Highlight Opportunities and Future Trends:**

We intend to identify potential benefits and advantages of Zero Trust that organizations can leverage to enhance their security posture. Additionally, we will explore emerging technologies and innovations that complement Zero Trust, paving the way for future trends in network security. By accomplishing these research objectives, we aim to contribute valuable insights to the ongoing discourse surrounding Zero Trust. This survey will serve as a resource for organizations seeking to make informed decisions about adopting Zero Trust as a proactive security strategy. Moreover, it will aid policymakers and decision-makers in understanding the significance of Zero Trust in countering evolving cyber threats and strengthening their organization's resilience in the face of dynamic challenges.

## 4. Related Work

Zero Trust has garnered significant attention in cybersecurity, leading to an extensive body of literature exploring its principles, applications, and effectiveness. This section presents a concise overview of related work conducted in Zero Trust, highlighting key findings and contributions from previous studies. Several research studies and publications have contributed to the understanding and advancement of the Zero Trust paradigm in the field of network security. These works have explored various aspects of Zero Trust, ranging from its theoretical foundations to its practical implementations, and have shed light on its applications, challenges, and potential opportunities. In the study, by (Yan and Wang, 2020) the researchers presented a comprehensive overview of the core principles of Zero Trust and its underlying philosophy. They highlighted the significance of adopting a zero-assumption approach, continuously verifying entities and the need for a fundamental shift in security strategies. This foundational work has provided the groundwork for subsequent research in the field. In a practical case study conducted by (Samaniego and Deters, 2018), the authors delved into the real-world implementation of Zero Trust in a financial institution. The study showcased the step-by-step process of transitioning from a traditional perimeter-based security model to a Zero Trust architecture. By analyzing the challenges faced during the deployment and the corresponding solutions, the researchers provided valuable insights for organizations planning to embark on a similar transformation journey.

The work of (Chen et al., 2021) explored the role of Zero Trust in securing Internet of Things (IoT) devices and its impact on enhancing the overall cybersecurity posture. Through simulations and real-world experiments, the researchers demonstrated the effectiveness of Zero Trust in preventing unauthorized access to IoT devices, mitigating potential threats, and protecting sensitive data. This study highlights the relevance of Zero Trust in the context of emerging technologies and the ever-expanding IoT ecosystem. Addressing the intersection of Zero Trust and regulatory compliance, (Tyler and Viana, 2021) analyzed how Zero Trust aligns with data protection laws, such as GDPR and HIPAA. By assessing the implications of Zero Trust on meeting compliance requirements, the researchers emphasized the significance of integrating security measures with regulatory frameworks, ensuring a robust and compliant security strategy. A comparative study by (Buck et al., 2021b) evaluated the effectiveness of Zero Trust against traditional perimeter-based security models. Through simulations and attack scenarios, the researchers demonstrated that Zero Trust significantly reduced the attack surface and effectively thwarted various cyber threats. The findings of this study underscore the superiority of the Zero Trust approach in enhancing network security. Another line of research (Anil, 2021) examined the impact of Zero Trust on organizational culture and employee behaviours. By exploring the challenges and opportunities of fostering a security-centric culture within organizations, the researchers highlighted the importance of addressing the human aspect of security alongside technological implementations. In recent years, several standard bodies, including NIST, have developed Zero Trust frameworks and guidelines. (Syed et al., 2022) reviewed these frameworks, evaluated their effectiveness, and further proposed enhancements to refine Zero Trust implementation practices. This work provides valuable insights for organizations seeking to adopt standardized approaches to Zero Trust deployment.

Through this survey, we aim to build upon the existing body of knowledge by consolidating and comparing these related works. By analyzing and synthesizing the findings from multiple studies, we strive to offer a comprehensive understanding of Zero Trust's applications, challenges, and opportunities in modern network security. The insights gained from this analysis will contribute to the broader knowledge base, inform current practices, and inspire future innovations in the realm of Zero Trust-based network security.

## 5. Zero Trust Implementation and Applications:

The implementation of Zero Trust represents a transformative shift in the approach to network security, where trust is no longer assumed, and access privileges are granted on a need-to-know basis. This section explores the practical aspects of implementing Zero Trust and examines its applications across diverse domains.

5.1. **Implementing Zero Trust**

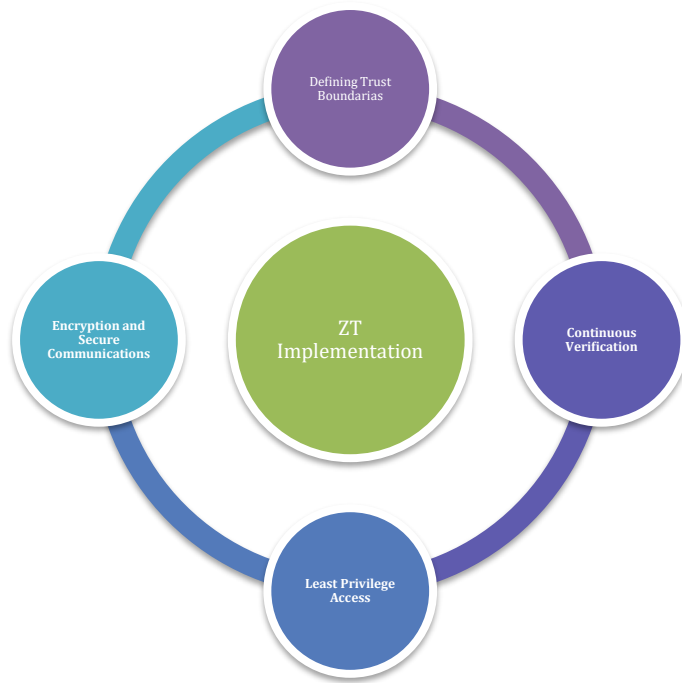

Figure 1. Four steps to zero trust implementation

### 5.1.1. Defining Trust Boundaries

The first step in implementing Zero Trust involves defining trust boundaries within the network. Organizations must identify critical assets, sensitive data, and access points and segment the network into smaller, isolated zones. Each zone acts as a virtual micro-perimeter, preventing unauthorized lateral movement and reducing the attack surface.

### 5.1.2. Continuous Verification

Zero Trust enforces a continuous verification process for all entities seeking access to network resources. This includes users, devices, applications, and even IoT devices. Multi-Factor Authentication (MFA), device health checks, and user behaviour analytics are used to ensure that only legitimate and authorized entities gain access.

### 5.1.3. Least Privilege Access

Zero Trust adheres to the principle of least privilege, where entities are granted only the minimum access required for their specific tasks. This fine-grained access control ensures that attackers have limited privileges and are contained within their designated zones even if a breach occurs.

### 5.1.4. Encryption and Secure Communications

Zero Trust emphasizes secure communication between entities. Encryption is employed for data in transit and at rest, adding an extra layer of protection against eavesdropping and data tampering.

## 5.2. Applications of Zero Trust

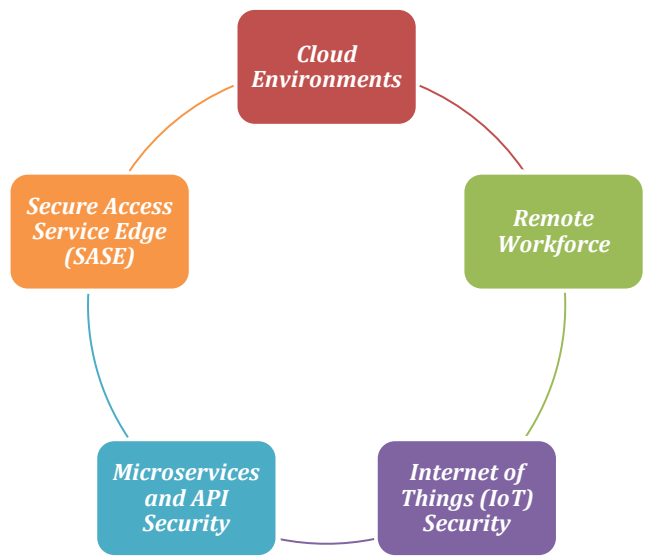

Figure 2. Zero Trust Application

- **Cloud Environments**

Zero Trust offers a robust security framework for protecting cloud-based assets and data with the increasing adoption of cloud services. By enforcing continuous verification and access controls, Zero Trust mitigates the risks associated with cloud security and provides granular control over cloud resources.

- **Remote Workforce**

Zero Trust is particularly well-suited for securing remote workers accessing corporate resources outside the traditional network perimeter. It ensures that remote devices and connections are continuously verified, reducing the risk of unauthorized access.

- **Internet of Things (IoT) Security**

Zero Trust plays a crucial role in securing interconnected devices as the IoT ecosystem expands. By enforcing strict access controls and continuous verification of IoT devices, Zero Trust prevents unauthorized access and mitigates the risk of IoT-related cyber threats. The implementation of a Zero Trust security model for Internet of Things (IoT) devices necessitates conducting a comprehensive inventory of all IoT devices present within the network. This inventory should entail categorizing each device based on its type, function, level of data sensitivity, and other qualities. The network ought to be designed in a manner that incorporates segmented zones and micro-perimeters, taking into account device function and sensitivity. imposing stringent restrictions on lateral movement between these zones is crucial. It is imperative to enforce continuous, granular authentication and authorization protocols for both user and device access. This entails confirming the identity context and permissions of individuals before providing them access to resources. It is necessary to monitor all activities and communications of Internet of Things (IoT) devices in order to develop behavior profiles and detect anomalies. This can be achieved through the utilization of analytics and artificial intelligence (AI) techniques, which enable the ranking of risks associated with IoT device behavior. It is recommended that an ecosystem adopts a default-deny access posture, wherein access is explicitly granted only after identity verification, and limited to the minimal necessary level. In order to address vulnerabilities, it is imperative to swiftly deploy security patches and firmware updates on Internet of Things (IoT) devices, with automation being employed whenever feasible. It is imperative that all network communication exchanged across Internet of Things (IoT) devices, gateways, and other endpoints be subjected to encryption, which should be accompanied by mutual authentication.

**Microservices and API Security:**

In modern application architectures, microservices and APIs (Application Programming Interfaces) are integral components. Zero Trust principles can be applied to secure the communication between microservices and ensure that authorized entities access only APIs.

- **Secure Access Service Edge (SASE):**

Zero Trust aligns closely with the SASE framework, which integrates network and security functions into a unified cloud-based service. Zero Trust principles form the foundation for securing access to applications and data in a distributedcloud-centric , network environment. Organizations can significantly enhance their network security posture by effectively implementing Zero Trust, adapt to dynamic threat landscapes, and protect critical assets from advanced cyber threats. By extending the applications of Zero Trust to various domains, organizations can create a resilient security framework that not only safeguards data and resources but also fosters trust and confidence in their digital operations.

## 6.  Challenges in Implementing Zero Trust:

Implementing the Zero Trust paradigm presents organizations with a spectrum of challenges that require thoughtful consideration and strategic mitigation. While the benefits of Zero Trust are compelling, navigating these challenges is integral to achieving successful adoption and realizing its full potential. This section elucidates the multifaceted challenges inherent in implementing Zero Trust and offers insights into addressing these complexities.

A. **Cultural Barriers and Organizational Resistance:**

Transitioning from traditional security models to Zero Trust necessitates a cultural shift within organizations. Employees accustomed to a perimeter-based trust model may encounter resistance to the continuous verification and least privilege access principles of Zero Trust. Overcoming this resistance demands effective change management strategies, comprehensive training, and transparent communication to engender a collective understanding of the paradigm shift.

B. **Technical Complexity and Integration:**

Integrating Zero Trust within an organization's infrastructure can be intricate and resource-intensive. Legacy systems, disparate technologies, and diverse platforms may pose compatibility challenges requiring meticulous planning and a phased implementation approach. Interconnecting various components while ensuring seamless functionality is pivotal to realizing the holistic benefits of Zero Trust.

C. **Balancing Security and Usability:**

Striking a delicate equilibrium between stringent security measures and user experience is a critical challenge in Zero Trust adoption. The constant verification and access controls may lead to potential user frustration and operational inefficiencies. Mitigating this challenge involves designing intuitive interfaces, streamlining authentication processes, and optimizing workflows to ensure that security enhancements do not compromise user productivity.

D. **Financial Considerations:**

The financial implications of Zero Trust implementation encompass investments in technology, personnel, and ongoing maintenance. Organizations must allocate resources for security infrastructure upgrades, training, and monitoring tools. Balancing these financial commitments against the anticipated benefits and risk reduction necessitates a thorough cost-benefit analysis and strategic resource allocation.

E. **Data Privacy and Regulatory Compliance:**

Continuous verification and access controls inherent to Zero Trust raise data privacy concerns and must align with regulatory frameworks. Ensuring compliance with data protection regulations, such as GDPR or HIPAA, while implementing robust security measures requires a nuanced understanding of legal requirements and meticulous data governance practices.

F. **Organizational Scalability:**

Zero Trust implementation strategies must accommodate organizational growth and scalability. Ensuring that Zero Trust architecture remains effective and efficient as the organization evolves necessitates careful architectural planning and adaptability to changing requirements.

In conclusion, the challenges associated with implementing Zero Trust underscore the necessity of a holistic and strategic approach. By addressing cultural, technical, usability, financial, and regulatory considerations, organizations can navigate these challenges and leverage the transformative potential of Zero Trust. The successful resolution of these hurdles empowers organizations to fortify their security posture and embrace a future where Zero Trust stands as a resilient and dynamic safeguard against contemporary cyber threats.

## 7. Security and Privacy Concerns:

While the Zero Trust paradigm offers a robust framework for enhancing network security, its implementation raises essential security and privacy considerations that warrant careful examination. As organizations adopt Zero Trust principles to mitigate cyber threats, assessing the potential risks and safeguards associated with this approach becomes imperative. This section delves into the security and privacy concerns that emerge within the context of Zero Trust, highlighting the need for a balanced and comprehensive approach (Campbell, 2020).

A. **Increased Attack Surface:**

The rigorous access controls and continuous verification inherent to Zero Trust can inadvertently expand the attack surface in certain scenarios. The intricate interplay of numerous components and verification mechanisms may introduce vulnerabilities that malicious actors could exploit. Organizations must diligently evaluate and fortify the various points of entry within the Zero Trust architecture to preempt potential breaches (Teerakanok et al., 2021b).

B. **Insider Threats and Credential Compromise:**

While focused on external threats, Zero Trust must also contend with insider threats. The stringent access controls may inadvertently amplify the impact of compromised credentials, as unauthorized users might find it more challenging to move laterally. Mitigating this concern necessitates robust user authentication mechanisms, real-time anomaly detection, and a vigilant approach to user behaviour analytics (Teerakanok et al., 2021c).

C. **Over-Reliance on Technology:**

Zero Trust heavily relies on technology for its efficacy, making the architecture susceptible to technological failures and vulnerabilities. More than relying on specific verification methods or technologies could lead to a single point of failure. Organizations must adopt a diversified and multi-faceted approach to verification, incorporating various technologies and strategies to enhance resilience.

D. **Privacy Implications:**

Continuous verification and real-time monitoring integral to Zero Trust may encroach upon user privacy. The collection and analysis of user data raise concerns about surveillance and potential misuse. Striking a balance between robust security measures and preserving user privacy requires implementing anonymization, pseudonymization, and strong data protection practices.

E. **Complexity of Monitoring and Management:**

The intricate nature of Zero Trust architectures, with their numerous components and interconnected verification processes, can lead to increased complexity in monitoring and management. Organizations must invest in comprehensive monitoring tools and skilled personnel to ensure the effective oversight of the entire Zero Trust ecosystem.

F. **Regulatory Compliance:**

The implementation of Zero Trust intersects with regulatory frameworks that govern data protection and privacy. Organizations must navigate the intricacies of compliance with regulations such as GDPR, HIPAA, and industry-specific standards. Adhering to these mandates while upholding the principles of Zero Trust necessitates a meticulous understanding of legal requirements and proactive measures to ensure alignment.

Given these security and privacy concerns, organizations must approach Zero Trust adoption with a holistic perspective that encompasses technical fortification, risk assessment, user education, and robust privacy safeguards. By recognizing and addressing these challenges, organizations can embrace Zero Trust while maintaining a vigilant commitment to security and privacy, ultimately fostering a cyber environment that harmonizes protection with respect for individual rights.

## 8. Opportunities and Future Trends

In cybersecurity, the landscape is perpetually evolving, necessitating a proactive approach to anticipate and address emerging challenges. The Zero Trust paradigm, with its foundational principles of continuous verification and dynamic access controls, addresses present vulnerabilities and provides a robust framework for harnessing opportunities and shaping future trends. This section elucidates the potential benefits of Zero Trust implementation and delves into the trajectory of its evolution within the cybersecurity domain.

### A. Advantages of Zero Trust:

Adopting Zero Trust engenders multifaceted advantages that resonate across diverse dimensions of network security and operational resilience. First and foremost, Zero Trust reduces the attack surface by enforcing rigorous access controls and segmenting the network into discrete zones. This pivotal measure mitigates thelateral movement of threats and confines potential breaches, thereby fortifying the security posture of organizations. Furthermore, Zero Trust enhances adaptability in the face of evolving cyber threats. The continuous verification of entities, including their users, devices, or applications, ensures a dynamic response to potential risks. By eschewing the traditional perimeter-based model, Zero Trust bolsters an organization's ability to thwart conventional and advanced attacks, engendering a heightened threat detection and response level. Moreover, the deployment of Zero Trust resonates harmoniously with contemporary technology paradigms, such as cloud computing and Internet of Things (IoT) ecosystems. The adaptable nature of Zero Trust facilitates the secure integration of these transformative technologies, fostering an environment conducive to innovation without compromising security.

### B. Technologies and Innovations:

The future trajectory of Zero Trust is intrinsically intertwined with the evolution of cutting-edge technologies. Artificial intelligence (AI) and machine learning (ML) hold the promise of enhancing Zero Trust's capabilities in anomaly detection, behavior analysis, and real-time threat identification. These technologies can augment the contextual understanding of user and device interactions, bolstering the accuracy of access decisions. Software-Defined Wide Area Networks (SD-WANs) and Secure Access Service Edge (SASE) frameworks are poised to synergize with Zero Trust principles, creating a harmonious fusion of network security and performance optimization. As organizations embrace these transformative network architectures, Zero Trust's role in orchestrating secure, seamless, and efficient communication channels becomes increasingly pivotal. Furthermore, the evolution of cryptographic protocols and encryption techniques contributes to the maturation of Zero Trust implementations. Integrating post-quantum cryptography and decentralized identity models aligns with Zero Trust's ethos of safeguarding data integrity and ensuring the authenticity of entities.

C. **Envisioning the Future of Zero Trust:**

As Zero Trust permeates the cybersecurity discourse, its adoption is anticipated to transcend industry boundaries and assume a foundational role in network security strategies. Organizations will embrace Zero Trust not merely as a reactionary measure but as a proactive stance integral to their risk management frameworks. The amalgamation of Zero Trust with cognitive computing ushers in a future where adaptive security becomes an inherent attribute of digital interactions. The innate ability of Zero Trust to discern context and intent, coupled with AI-driven decision-making, culminates in a security fabric that instinctively evolves in response to threats. In conclusion, the opportunities and future trends surrounding Zero Trust are imbued with transformative potential. Its capacity to fortify security landscapes, harmonize with emerging technologies and envision a dynamic cybersecurity ecosystem substantiates Zero Trust's position as a vanguard of network protection. By embracing Zero Trust and charting a trajectory that aligns with its principles, organizations are poised to navigate the complex cyber terrain with resilience, agility, and unwavering trustworthiness.

## 9. Conclusion

The Zero Trust paradigm offers a transformative approach to modern cybersecurity, challenging traditional models and providing a forward-looking solution. This survey has delved into its principles, practical applications, challenges, and future potential. Zero Trust's core principles of continuous verification and least privilege access align with cloud security, remote work, and IoT protection demands. While challenges like cultural shifts and technical complexities exist, they can be overcome to leverage Zero Trust's benefits fully. Integrating Zero Trust with emerging technologies like AI, machine learning, and SD-WAN promises enhanced efficacy. This synergy envisions a responsive security landscape ready to counter sophisticated cyber threats. In summary, embracing Zero Trust empowers stakeholders to navigate the evolving cybersecurity landscape with resilience and adaptability. By redefining trust and implementing dynamic access controls, Zero Trust stands as a pillar of network security, contributing to a safer digital realm.

## 10. Reference

Alagappan A, Venkatachary SK, Andrews LJB. Augmenting Zero Trust Network Architecture to enhance security in virtual power plants. Energy Reports 2022;8:1309–20. https://doi.org/10.1016/J.EGYR.2021.11.272.

Alevizos L, Ta VT, Hashem Eiza M. Augmenting zero trust architecture to endpoints using blockchain: A state-of-the-art review . SECURITY AND PRIVACY 2022;5. https://doi.org/10.1002/SPY2.191.

Anil G. A Zero-Trust Security Framework for Granular Insight on Blind Spot and Comprehensive Device Protection in the Enterprise of Internet of Things ( E- IOT ). Department of Computer Science and Engineering, 2021:1–25.

Astillo PV, Choudhary G, Duguma DG, Kim J, You I. TrMAps: Trust Management in Specification-Based Misbehavior Detection System for IMD-Enabled Artificial Pancreas System. IEEE J Biomed Health Inform 2021;25:3763–75. https://doi.org/10.1109/JBHI.2021.3063173.

Buck C, Olenberger C, Schweizer A, Völter F, Eymann T. Never trust, always verify: A multivocal literature review on current knowledge and research gaps of zero-trust. Comput Secur 2021a;110. https://doi.org/10.1016/J.COSE.2021.102436.

Buck C, Olenberger C, Schweizer A, Völter F, Eymann T. Never trust, always verify: A multivocal literature review on current knowledge and research gaps of zero-trust. Comput Secur 2021b;110. https://doi.org/10.1016/J.COSE.2021.102436.

Campbell M. Beyond Zero Trust: Trust Is a Vulnerability. Computer (Long Beach Calif) 2020;53:110–3. https://doi.org/10.1109/MC.2020.3011081.

Cao Y, Pokhrel SR, ZHU Y, Ram Mohan Doss R, Li G. Automation and Orchestration of Zero Trust Architecture:Potential Solutions and Challenges 2022. https://doi.org/10.36227/TECHRXIV.21385929.V1.

Chen Z, Yan L, Lü Z, Zhang Y, Guo Y, Liu W, et al. Research on Zero-trust Security Protection Technology of Power IoT based on Blockchain. J Phys Conf Ser 2021;1769. https://doi.org/10.1088/1742-6596/1769/1/012039.

Egerton H, Hammoudeh M, Unal D, Adebisi B. Applying Zero Trust Security Principles to Defence Mechanisms Against Data Exfiltration Attacks. Security and Privacy in the Internet of Things: Architectures, Techniques, and Applications 2021:57–89. https://doi.org/10.1002/9781119607755.CH3.

Gupta T, Choudhary G, Sharma V. A Survey on the Security of Pervasive Online Social Networks (POSNs) 2018. https://doi.org/10.22667/JISIS.2018.05.31.048.

Rahmadika S, Astillo PV, Choudhary G, Duguma DG, Sharma V, You I. Blockchain-Based Privacy Preservation Scheme for Misbehavior Detection in Lightweight IoMT Devices. IEEE J Biomed Health Inform 2023;27:710–21. https://doi.org/10.1109/JBHI.2022.3187037.

Ramezanpour K, Jagannath J. Intelligent zero trust architecture for 5G/6G networks: Principles, challenges, and the role of machine learning in the context of O-RAN. Computer Networks 2022;217. https://doi.org/10.1016/j.comnet.2022.109358.

Ray PP. Web3: A comprehensive review on background, technologies, applications, zero-trust architectures, challenges and future directions. Internet of Things and Cyber-Physical Systems 2023;3:213–48. https://doi.org/10.1016/J.IOTCPS.2023.05.003.

Samaniego M, Deters R. Zero-trust hierarchical management in IoT. Proceedings - 2018 IEEE International Congress on Internet of Things, ICIOT 2018 - Part of the 2018 IEEE World Congress on Services 2018:88–95. https://doi.org/10.1109/ICIOT.2018.00019.

Syed NF, Shah SW, Shaghaghi A, Anwar A, Baig Z, Doss R. Zero Trust Architecture (ZTA): A Comprehensive Survey. IEEE Access 2022;10:57143–79. https://doi.org/10.1109/ACCESS.2022.3174679.

Teerakanok S, Uehara T, Inomata A. Migrating to Zero Trust Architecture: Reviews and Challenges. Security and Communication Networks 2021a;2021. https://doi.org/10.1155/2021/9947347.

Teerakanok S, Uehara T, Inomata A. Migrating to Zero Trust Architecture: Reviews and Challenges. Security and Communication Networks 2021b;2021. https://doi.org/10.1155/2021/9947347.

Teerakanok S, Uehara T, Inomata A. Migrating to Zero Trust Architecture: Reviews and Challenges. Security and Communication Networks 2021c;2021. https://doi.org/10.1155/2021/9947347.

Tyler D, Viana T. Trust no one? A framework for assisting healthcare organisations in transitioning to a zero-trust network architecture. Applied Sciences (Switzerland) 2021;11. https://doi.org/10.3390/APP11167499.

Velumani R, Sudalaimuthu H, Choudhary G, Bama S, Jose MV, Dragoni N. Secured Secret Sharing of QR Codes Based on Nonnegative Matrix Factorization and Regularized Super Resolution Convolutional Neural Network. Sensors 2022;22. https://doi.org/10.3390/S22082959.

Yan X, Wang H. Survey on Zero-Trust Network Security. Communications in Computer and Information Science 2020;1252 CCIS:50–60. https://doi.org/10.1007/978-981-15-8083-3_5.